\documentclass[fleqn,usenatbib]{mnras}

\usepackage{newtxtext,newtxmath}

\usepackage[T1]{fontenc}
\usepackage{ae,aecompl}

\usepackage{graphicx}	
\usepackage{amsmath}	
\usepackage{amssymb}	
\usepackage{tabularx}
\usepackage{caption}
\usepackage[flushleft]{threeparttable}

\title[NGC 2903's bulge, bar, and outer disc]{The VIRUS-P Exploration of Nearby Galaxies (VENGA): The stellar populations and assembly of NGC 2903's bulge, bar, and outer disc}

\author[A. Carrillo et al.]{Andreia Carrillo$^{1,2}$,\thanks{E-mail: andreiac@utexas.edu}
Shardha Jogee$^{1}$,
Niv Drory$^{3}$,
Kyle F. Kaplan$^{4}$,
Guillermo Blanc$^{5,6}$,
\newauthor Tim Weinzirl$^{7}$,
Mimi Song$^{8}$,
Rongxin Luo$^{9}$
\\
\\
$^{1}$Department of Astronomy, University of Texas at Austin, 2515 Speedway, Stop C1400, Austin, TX 78712-1205, USA\\
$^{2}$Large Synoptic Survey Telescope Corporation Data Science Fellow\\
$^{3}$McDonald Observatory, The University of Texas at Austin, 1 University Station, Austin, TX 78712, U.S.A.\\
$^{4}$SOFIA-USRA, NASA Ames Research Center, MS N232-12, Moffett Field, CA 94035-1000, USA\\
$^{5}$The Observatories of the Carnegie Institution for Science, 813 Santa Barbara St., Pasadena, CA 91101, USA\\
$^{6}$Departamento de Astronomia, Universidad de Chile, Castilla 36-D, Santiago, Chile\\
$^{7}$ School of Physics and Astronomy, The University of Nottingham, University Park, Nottingham, NG7 2RD, UK\\
$^{8}$Astrophysics Science Division, Goddard Space Flight Center, Code 665, Greenbelt, MD 20771, USA\\
$^{9}$Shanghai Astronomical Observatory, 80 Nandan Road, Shanghai 200030, China
}

\date{Accepted XXX. Received YYY; in original form ZZZ}

\pubyear{2018}

\begin{document}
\label{firstpage}
\pagerange{\pageref{firstpage}--\pageref{lastpage}}
\maketitle

\begin{abstract}
We study the stellar populations and assembly of the nearby spiral galaxy NGC 2903's bulge, bar, and outer disc using the VIRUS-P Exploration of Nearby Galaxies IFS survey. We observe NGC 2903 with a spatial resolution of 185 pc using the Mitchell Spectrograph's 4.25 arcsec fibres at the 2.7 Harlan J. Smith telescope. Bulge-bar-disc decomposition on  the 2MASS  $K_s$-band image of NGC 2903  shows that it has $\sim$ 6\%, 6\%, and 88\%, of its stellar mass in the bulge, bar, and outer disc, respectively, and its bulge has a low S\'ersic index of $\sim 0.27$, suggestive of a disky bulge.  We perform stellar population synthesis and find that the outer disc has 46\% of its mass in stars $>$ 5 Gyr, 48\% in stars between 1 and 5 Gyr, and $<$10\% in younger stars. Its stellar bar has 65\% of its mass in ages 1-5 Gyr  and has metallicities similar to the outer disc, suggestive of the evolutionary picture where the bar forms from disc material. Its bulge is mainly composed of old high-metallicity stars though it also has a small fraction of young stars. We find enhanced metallicity in the spiral arms and central region, tracing areas of high star formation as seen in the H$\alpha$ map. These results are consistent with the idea that galaxies of low bulge-to-total mass ratio and low bulge S\'ersic index like NGC 2903 has not had a recent major merger event, but has instead grown mostly through minor mergers and secular processes. 

\end{abstract}




\section{Introduction}
\label{sec:intro}
Observational constraints on the assembly history of different components of a galaxy are essential to understand how quickly and through what mechanisms galaxies grow. The outer disc, primary bar, and different types of  central bulges (classical, disky and boxy/peanut) are fundamental stellar components of spiral galaxies that  define the Hubble sequence \citep{kormendy93,athanassoula03,sheth05}. These components are thought to be built through different evolutionary pathways and their morphology, kinematics, star formation rate (SFR), stellar ages, stellar metallicities, and star formation history (SFH), provide important  constraints on the assembly history of these components and their host galaxy. 

Simulations have suggested that the outer disc grows from minor mergers or gas accretion through cold and dense filaments \citep{somervilledave15}. After which, bar-formation could be spontaneously or tidally induced \citep{romano08}. The bar then efficiently redistributes mass and angular momentum in the galaxy \citep{shlosman93,knapen95,athanassoula05,jogee05}  and drives gas inflows into the circumnuclear regions of the galaxy. 

The bulge, on the other hand, comes in different forms that have different proposed evolutionary scenarios. Classical bulges are dominated by random motion (characterized by velocity dispersion, $\sigma$) rather than ordered motion (characterized by circular velocity, v), therefore having low v/$\sigma$. A classical bulge is thought to be built generally by violent relaxation in dry or modestly gas-rich galaxy major mergers. Disky pseudobulges have high v/$\sigma$ and are dominated by ordered motion, reminiscent of galactic discs. They are believed to have been built via gas that is funneled to the central region of a galaxy \citep{kormendy93, KK04, jogee05, athanassoula05, weinzirl09}. Boxy (or peanut) pseudobulges form from stellar bar material scattered to larger scale heights via instabilities or vertical resonances \citep{combes90,athanassoula05}. It is important to note that a galaxy does not necessarily have just one type of bulge. 
For example, a disky pseudobulge can coexist with a boxy pseudobulge (e.g., \citealt{barentine12}) or a classical bulge (e.g., \citealt{erwin15}). 

High resolution images have allowed for morphological and structural studies that place some constraints on a galaxy's assembly history \citep{kormendy93,KK04,weinzirl09,weizirl14,fisher08, jogee04, marinova07, jogee09, weinzirl11}. Integral Field Spectroscopy (IFS or IFU) surveys at high spatial resolution can strongly complement morphological studies by constraining the stellar populations and ionized gas properties of different galaxy components. One such survey is the VIRUS-P Exploration of Nearby Galaxies (VENGA) IFU survey \citep{blanc13a} that includes observations of 30 nearby spirals from their central regions to their outskirts. VENGA complements many recent IFU surveys (MaNGA, \citealt{bundy15}; CALIFA, \citealt{sanchez12}; SAMI, \citealt{croom12}), that target large samples of galaxies out to large distances, have spatial resolution of typically several kpc, and  have primarily provided important statistical insights on global properties of galaxies. With VENGA, we can specifically explore  the assembly of the  outer disc, primary bar, and different types of central bulges.

We are doing a pilot study on the stellar populations and mass build-up of the bulge, the bar, and the outer disc for one of the galaxies in the VENGA survey, NGC 2903. It is an isolated spiral galaxy \citep{irwin09} with circumnuclear star formation, a bar, and grand-design spiral arms. With this pilot study, we aim to advance the field of stellar populations and aid our understanding of the evolution in the different parts of a galaxy because of the following advantages: (1) NGC 2903's low distance (8.9 Mpc) leads to a superb spatial resolution (185 pc), and the large coverage from the bulge to the outer disc allows for the study of spatially-resolved galaxy properties; (2) its structural properties are interesting: it is a SAB(rs)bc spiral \citep{devaucouleurs95} with a strong bar in the near-infrared (NIR ellipticity of 0.88 and a bar-to-total (Bar/T) light ratio = 0.06; see Table \ref{tab:galfit_decomp}), a low bulge-to-total (B/T) light ratio of 0.06,  and low S\'ersic  index, n$\sim$0.27 (see Table \ref{tab:galfit_decomp}). 
We aim to compare our stellar population synthesis results of NGC 2903 to the study by \citet{weinzirl09} which suggests, based on their comparison of a large sample of spirals with predictions from two semi-analytical models \citep{khochfar05,khochfar06,hopkins09}, that spiral galaxies with low bulge to total mass ratio (B/T< 0.2) have not undergone a major merger since z$\sim$2, and have built their outer discs as well as part of their bulge at later times through gas accretion, minor mergers, and secular processes. This is supported by more recent studies on bulge formation with hydrodynamical simulations (e.g. \citealt{gargiulo19,tacchella19}) that show the importance of secular evolution and significance of disc formation in building low S\'ersic index and low B/T bulges; (3) it is one of the barred galaxies in \citet{kaplan16} that also uses VENGA to look at gas-phase metallicity with 7 different indicators providing us with emission line analysis results to compare to; (4) it is a well-studied galaxy and has a wealth of ancillary data (e.g., 2MASS near-infrared among others) which we can take advantage of to do a morphological study in combination with the spectral analysis. 

This paper is organized as follows: In Section \ref{sec:methods} we describe observation, data reduction and data cube building, we outline in Section \ref{sec:galfit} the bulge-bar-disc decomposition using GALFIT, in Section \ref{sec:sps} we explain the stellar population fitting procedure, in Section \ref{sec:SFH} we present mass-build up histories for the bulge, bar, and outer disc, in Section \ref{sec:results} we discuss the implications of our derived stellar population maps and star formation histories, and lastly, we summarize our results in Section \ref{sec:conclusion}.

\section{VENGA Sample and Data Reduction}
\label{sec:methods}


\subsection{Observations}
\label{sec:vengaobs}

VENGA \citep{blanc13a} includes 30 nearby spirals observed over $\sim$150 nights with the Mitchell Spectrograph on the 2.7 meter Harlan J. Smith telescope \citep{gary08}. The survey benefits from high spatial sampling of typically a few 100 pc resolution (fibre size of 4.25$\arcsec$) with 80\% of the galaxies less than 20 Mpc away, large coverage from the bulge to the outer disc,  broad wavelength range (3600-6800 \AA), and medium spectral resolution (120 km/s at 5000 \AA~or R$\sim$1000).

\begin{figure}
\begin{center}
\includegraphics[width=0.4\textwidth]{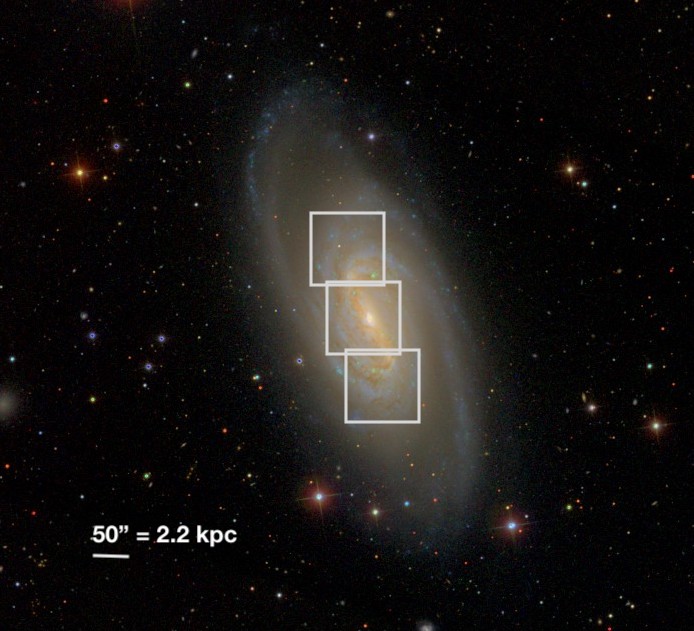}

\caption{SDSS image of NGC 2903 (inverted colormap) with 3 Mitchell Spectrograph pointings, each with a field of view of 110\arcsec $\times$ 110\arcsec. The VENGA pointings encompass the bulge, bar, and part of the outer disc of NGC 2903.}
\label{fig:sdss_image}
\end{center}
\end{figure}

Three pointings, each with three dithers were done for NGC 2903, with every dither observed at 4800s exposure time. The dithering compensates for the 1/3 filling factor of the spectrograph and the three pointings enables us to observe from the bulge to the outer disc with a field of view of 110\arcsec $\times$ 110\arcsec~per pointing as seen in Figure \ref{fig:sdss_image}. Observations were done with a blue setup (3600-5800 \AA) and a red setup (4600-6800 \AA) to increase the wavelength coverage of the survey without compromising the medium spectral resolution. These were then later combined during the reduction procedure. 

These observations result in data with high signal-to-noise ratio (SNR) of up to $\sim$300 in the central region. We have a spatial resolution of 185 pc at the distance of NGC 2903 (8.9 Mpc), enough to resolve the bulge area with multiple spaxels. 

\subsection{Data Reduction and Building Data Cube}
\label{sec:data_reduction} 

We refer the reader to \citet{blanc13a} for details on the VENGA data reduction. To summarize, we used the pipeline VACCINE \citep{adams11} which does bias and dark subtraction, flat fielding, wavelength calibration, cosmic ray rejection, and sky subtraction after which, the astrometry was determined and the data was flux-calibrated. This was a two-step process: first by using a standard star for relative flux calibration and next by using SDSS $g$ and $r$ band images, convolved to the fiber size of Mitchell spectrograph, for absolute flux-calibration and astrometry. Lastly all science frames were combined into a datacube (two versions for each galaxy with linearly-sampled or logarithmically-sampled wavelengths) that contains the wavelength, flux, flux error, spatial position, and instrumental resolution. 

Additionally, we Voronoi-binned \citep{cappellari03} the data to increase the SNR to at least 50 per bin. This had a net effect of combining spaxels on the edges of the field of view. In the end, we had a total of 6,812 spaxels available for full-spectrum fitting for NGC 2903.

\section{Bulge-bar-disc decomposition}
\label{sec:galfit}

We follow the methods \citet{weinzirl09} in decomposing the galaxy light into the bulge, bar, and disc components. This method specifically uses GALFIT \citep{peng10} to fit different S\'ersic profiles to the galaxy light. We use the 2MASS $K_s$-band image from  2MASS Large Galaxy Atlas \citep{jarrett03} for our decomposition because it best traces stellar mass and the stellar mass-to-light ratio in the $K_s$-band remains roughly constant with stellar population age. 
We also determined a PSF image from stars in the field, to be fed into GALFIT. 

\begin{figure*}
\begin{center}
\includegraphics[width=0.7\textwidth]{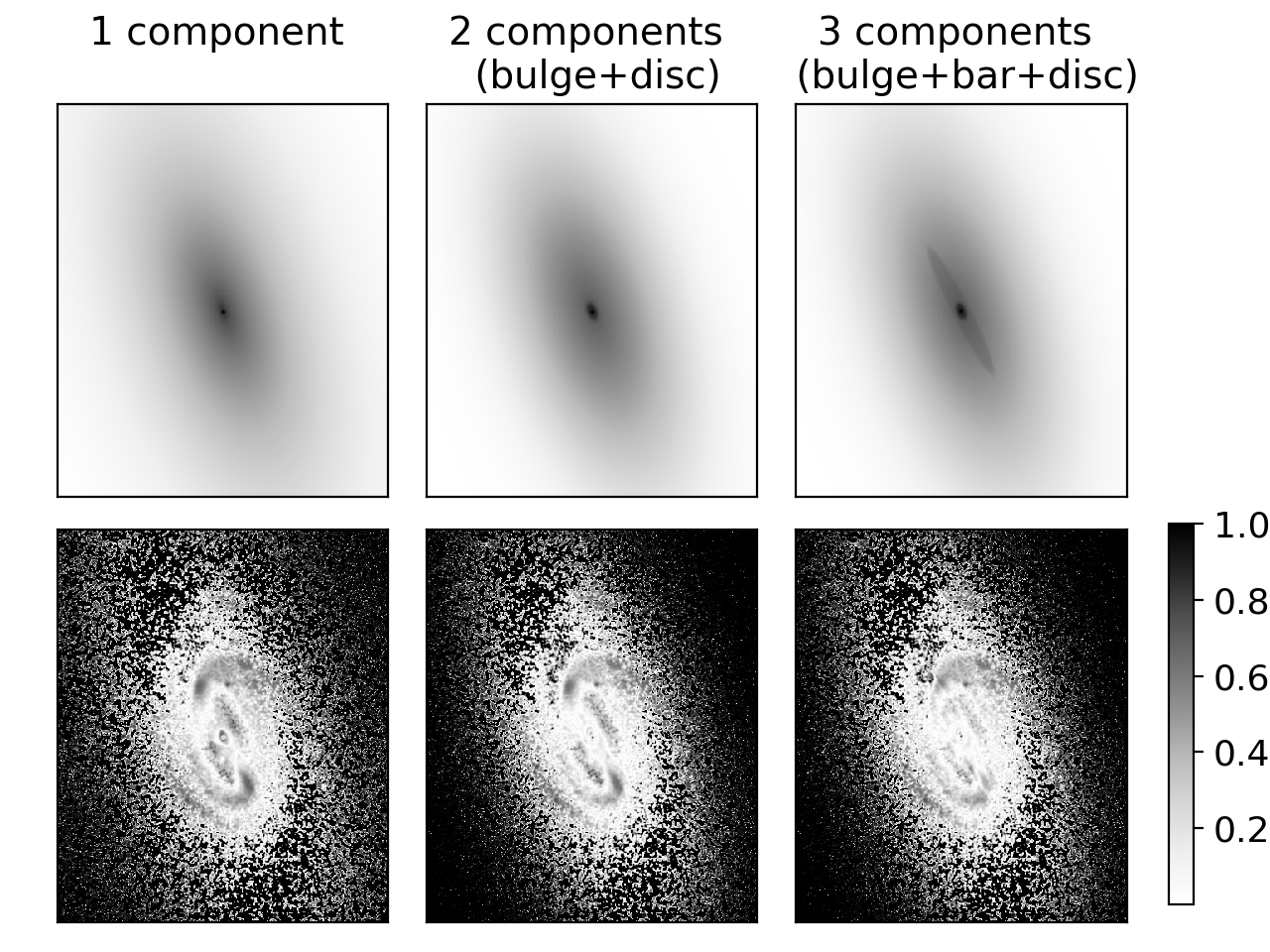}

\caption{GALFIT decomposition of NGC 2903 using 2MASS $K_S$-band image. The top row shows the best fit GALFIT model, and the bottom row shows the residual (data-model) divided by the data. The grayscale applies to the bottom row where lighter shade indicates smaller residual fraction and therefore a better fit. The columns are arranged as one component fit ($\chi^2$/DOF = 1.55), 2 component fit with the bulge and outer disc decomposition  ($\chi^2$/DOF = 1.17), and three component fit with the bulge, bar, and outer disc decomposition ($\chi^2$/DOF = 1.11) following \citet{weinzirl09}. The residuals for the 3-component decomposition shows that it best fits the light profile of NGC 2903.}
\label{fig:galfit}
\end{center}
\end{figure*}

We do the following steps based on \citet{weinzirl09}: we first fit with a single S\'ersic  component which determines the total luminosity, the center of the galaxy, and the inclination of the outer disc which are then held fixed for the later stages. Next we fit two components, a S\'ersic  component representative of the bulge, and an exponential disc. We then proceed to a three component fit with an exponential disc and two S\'ersic  components (for the bulge and for the bar) using the best fit model from the two component (bulge+disc) fit as initial guesses on the parameters. The models and the residuals divided by the data are shown in Figure \ref{fig:galfit}. We decide whether the 2-component (bulge-disc) or 3-component (bulge-bar-disc) decomposition is better based on the prescription from Section 3.3 in \citet{weinzirl09}.  An indication that the 3-component fit is most appropriate for NGC 2903 can be seen in the residual divided by the data map (see bottom row of Figure \ref{fig:galfit}), where lighter regions indicate lower residual over data and therefore a better fit. Since the galaxy has a bar as seen in Figure \ref{fig:sdss_image}, the residuals in the one component and two-component fits contain prominent and $m=2$ symmetric light distribution pattern due to the unsubtracted bar light. However, this residual pattern is significantly reduced with the three component (bulge+bar+disc) fit. Other light patterns in the residual map remain, though these are due to the spiral arms which are unaccounted for in our light decomposition. Table \ref{tab:galfit_decomp} shows the component-to-total ratios and statistical results from this GALFIT analysis of NGC 2903. 

\begin{table*}
\begin{threeparttable}
	\caption{Results of GALFIT decomposition on 2MASS $K_s$-band image}
	\label{tab:galfit_decomp}
	
    \begin{center}
	\begin{tabular}{c c c c c c c c c}
		\hline
        FIT & & $\chi^2$ & $\chi^2$/DOF & $R_e$ or h (\arcsec, kpc) & S\'ersic  index n & b/a & PA & X/T \\
        (1) & (2) & (3) & (4) & (5) & (6) & (7) & (8) & (9) \\        
        \hline
        \textbf{1-Component} & S\'ersic  & 294,523 & 1.55 & 173.51, 7.55 & 1.790$\pm$0.005 & 0.449$\pm$0.001 & 17.58 & 1.000 \\
        \hline
        \textbf{2-Component} & Bulge &  221,773 & 1.17 & 13.94, 0.61 & 0.390$\pm$0.006 & 0.533$\pm$0.002 & 20.39  & 0.0644 $\pm$ 0.0001 \\
         & Disc &  & & 98.93, 4.31 & 1.00 & 0.449 & 17.58  & 0.9356 $\pm$ 0.0002 \\ 
         \hline
        \textbf{3-Component} & Bulge &   202,364 & 1.11 & 13.25, 0.58 & 0.266$\pm$0.005 & 0.529$\pm$0.002 & 19.79  & 0.0571 $\pm$ 0.0001 \\
         & Disc &  & & 104.60, 4.55 & 1.00 & 0.449 & 17.58  & 0.8835 $\pm$ 0.0004 \\ 
         & Bar &  & & 116.62, 5.07 & 0.060$\pm$0.002 & 0.124$\pm$0.001 & 26.96 &  0.0593 $\pm$ 0.0003\\ 
		\hline         
		\end{tabular}
	\begin{tablenotes}
      \small
      \item Notes: (1) 3-Stage GALFIT fitting with 1, 2, and 3 components. (2) GALFIT light profile/galaxy component being fit (3) $\chi^2$ for each step (4) reduced $\chi^2$ (5) Effective radius or scale length in case of the disc measured in arcseconds (6) S\'ersic  index n \citep{sersic63} (7) semi major axis to semi minor axis ratio i.e. 1 - ellipticity, $e$ (8) Position angle measured in degrees (9) component to total light ratio.
    \end{tablenotes}
    \end{center}
\end{threeparttable}
\end{table*}

\begin{figure}
\begin{center}
\includegraphics[width=0.5\textwidth]{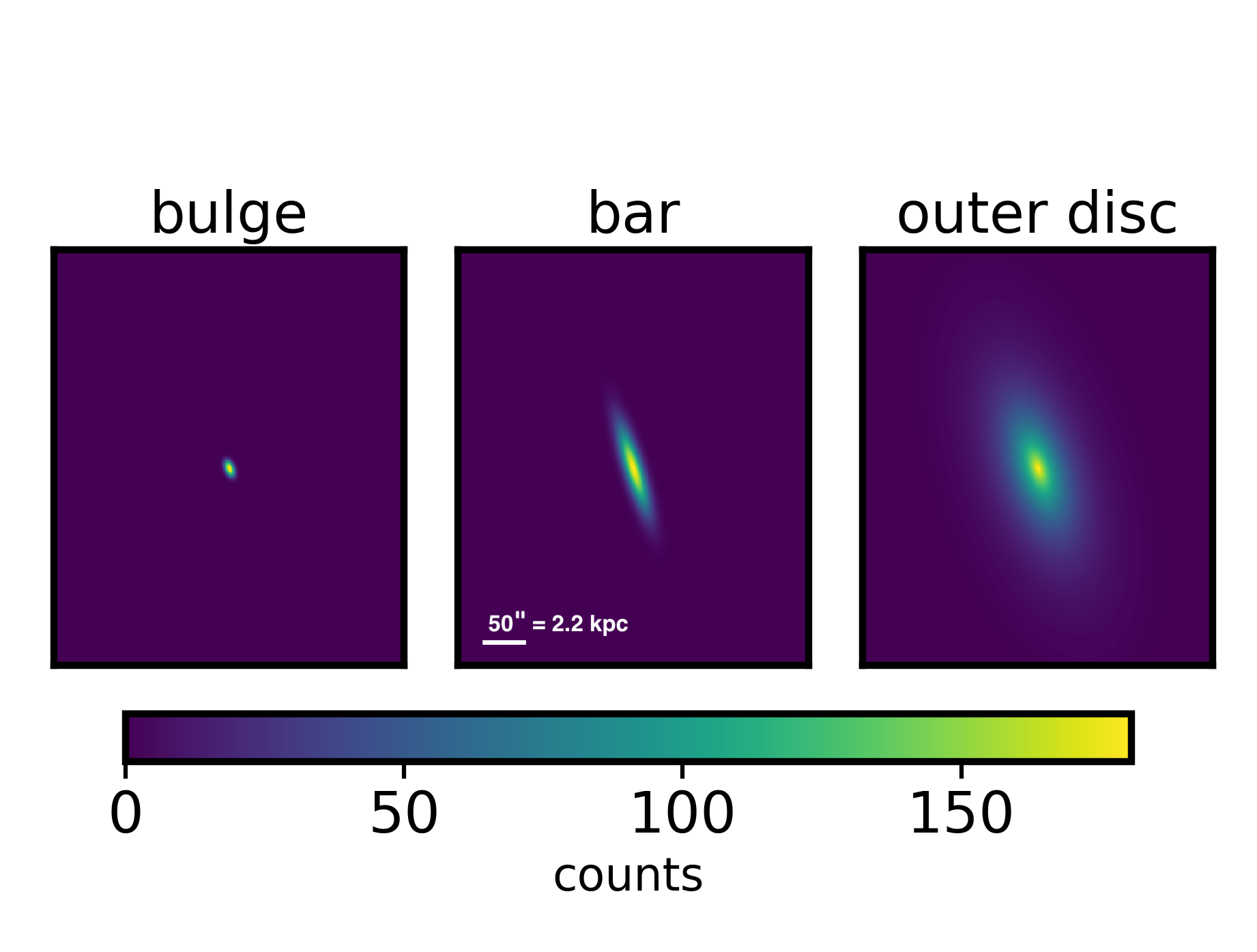}

\caption{Light contributions from the bulge(left), bar(center), and disc(right) from the 3-component GALFIT decomposition. The colorbar indicates the brightness counts. }
\label{fig:subcomponents}
\end{center}
\end{figure}

\begin{figure}
\begin{center}
\includegraphics[width=0.3\textwidth]{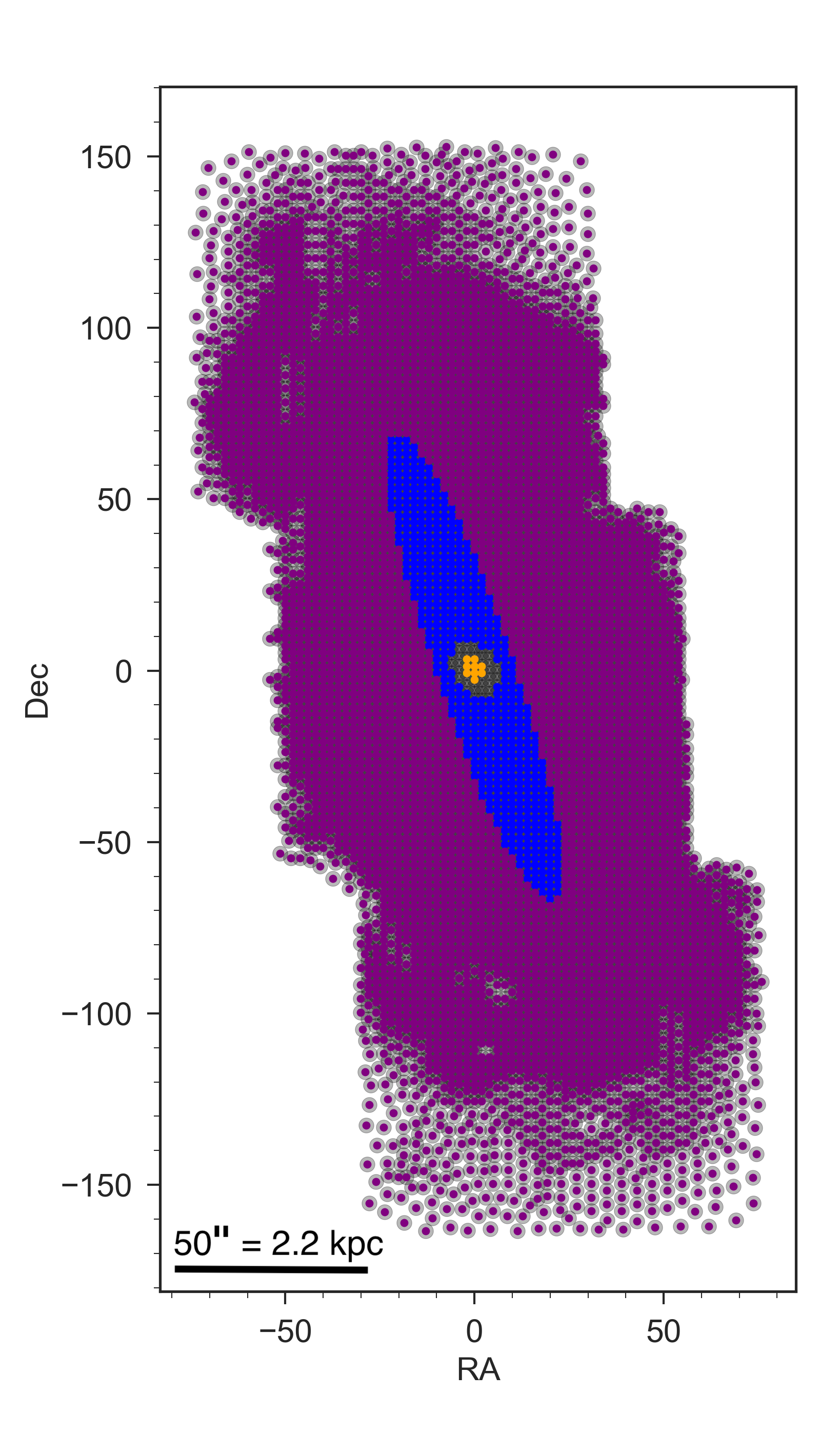}

\caption{We assign spaxels to the bulge, bar, and outer disc if these components contribute to $>$ 50 \% of the total light  at each location. The spaxels  are shown as  orange for the bulge, blue for the bar, and purple for the outer disc. The grey areas have spaxels with no dominating (i.e. $>$50\%) light source. They were therefore not included in calculating the mass build up history for the individual galactic components. }
\label{fig:spaxels}
\end{center}
\end{figure}

We then fed the 3-component best-fit model back to GALFIT to get the image blocks for each light-component (see Figure \ref{fig:subcomponents}). For a given position, we now know how much light is contributed by the bulge, bar, and disc components. We assign the NGC 2903 VENGA spaxels by matching their position to the GALFIT decomposition and determining which component contributes > 50\% of the light for that area. This breakdown of spaxel assignment is illustrated in Figure \ref{fig:spaxels} where nine bulge spaxels are shown in orange, 640 bar spaxels are in blue, 6,062 outer disc spaxels are in purple, and 101 unassigned spaxels are in gray. Though we know how much each component contributes to the light, we do not know the breakdown of the stellar populations that should be ascribed to each component at a given spaxel. We therefore chose to assign a spaxel to one component as long as that component dominates more than half of the light in that region. Going forward, we use these spaxel memberships to assign a given spaxel to the bulge or bar or outer disc components, to determine their associated properties. The unassigned spaxels (1.5\%) were excluded from the individual component mass build-up histories. We do not use a more stringent spaxel assignment (i.e. $>$70\%) as this yields no spaxels for the bulge. This also decreases the number of bar spaxels to 97 and outer disc spaxels to 5,638.

\section{Stellar Populations Analysis}
\label{sec:sps}

Fitting physically-motivated combinations of stellar populations to IFU data is challenging, and so parameterization of the SFH is usually implemented (see discussions in \citealt{conroy13}). However, numerous codes do exist for fitting stellar populations to IFU data (STARLIGHT: \citealt{cidfernandes05}, PIPE3D: \citealt{sanchez16}, FIREFLY: \citealt{wilkinson17}) that have nonparametric SFH. The advantage of a nonparametric solution is that it does not assume a declining, rising, or constant SFH, and is therefore a better approach in getting the SFH of a spiral galaxy, which may have a bursty or stochastic star formation history. 

For fitting the stellar populations in this study, we used the \textit{Penalized Pixel-Fitting} (pPXF) method \citep{cappellari04, cappellari17}. Though usually employed for getting gas and stellar kinematics, it has also been used in a number of stellar populations studies (e.g., \citealt{onodera12,mcdermid15}). In general, pPXF derives the mass-weighted, best-fit template as a linear combination of stellar populations with different ages and metallicities. Regularization is suggested for using pPXF in stellar population studies as deriving the SFH from a grid of templates is an ill-conditioned inverse problem (see discussion in Section 3.5 from \citealt{cappellari17}) and requires further assumptions. This is especially important for low SNR data like other IFU surveys of nearby galaxies where over-fitting to noisy data could be an issue. A regularized solution gives the smoothest star formation history consistent with the data. However, with our high SNR data, the non-regularized best fit model from pPXF performed well in our cross-validation tests and produced similar SFH trends to the regularized case (tested on 120 spaxels), as we collapse our age resolution into four bins in the end. The differences in the average ages and metallicities between the non-regularized and regularized fits are also quite small i.e. $0.50 \pm 1.9$ Gyr for age and $0.13 \pm 0.06$ dex for metallicity. From these tests, we decided to proceed on using pPXF without regularization which benefits from being computationally less expensive while still producing physically-motivated results.


Throughout the full-spectrum fitting process, we mask out wavelength ranges that have poor sky subtraction. We also mask higher order Balmer lines (up to H$\theta$) to improve the fit, especially for areas with high star formation activity. In determining the best fit combination of stellar populations, we normalize the continuum with a Legendre polynomial (LP) 
Later on however, we do derive the reddening for each spaxel from the continuum (see Section \ref{sec:dust}) in order to get the intrinsic luminosity and therefore the intrinsic mass of the galaxy.

\subsection{Fine-tuning of the template library}
\label{sec:library}

We first used stellar population templates from the MILES template library from \citet{vazdekis10} that were already provided within the pPXF package distribution. Specifically, these spectra are based on the MILES empirical stellar library with a \citet{salpeter55} initial mass function (IMF), and Padova isochrone tracks \citep{girardi00}. The spectra have FWHM of 2.51 \AA~that cover 6 metallicities  from [M/H]~=~-2.3 to [M/H]~=~+0.22 and a full age range (from 63.1 Myr to 15.85 Gyr). We convolved the templates with the wavelength-dependent instrumental dispersion of our data. 
 
However, using these templates for NGC 2903, a star-forming galaxy, forced the best fit spectra to contain a sizable portion of the mass to come from the youngest template at the lowest metallicity. This is unphysical unless pristine, un-enriched gas is funneled into the galaxy. The more likely scenario is that young stars are present and pPXF erroneously tries to fit the bluest template, which in MILES correspond to the youngest, most metal-poor templates.

To alleviate this problem, we used the more extensive library Base GM (Rosa Gonzalez-Delgado priv. comm.) adopted from the code STARLIGHT. This template library also includes the \citet{vazdekis10} single stellar populations (SSP) with the addition of younger templates from \citet{gonzalezdelgado05}. These younger models consist of 15 ages from 1 to 63 Myr at four metallicities from [M/H] = -0.71 to [M/H] = + 0.22, and uses Geneva isochrone tracks \citep{Schaller92,Schaerer93,Charbonnel93} as well as a Salpeter IMF. 

Base GM templates has a FWHM of 6 \AA, which matches the CALIFA resolution \citep{sanchez16} and is higher or lower than our data's spectral resolution (4.6\AA ~to 6.1\AA), depending on the wavelength range under consideration. pPXF is ran with the prescription that the template library has similar or better resolution than the observed data. We therefore had to degrade our data to the same resolution as the templates and re-measure the final FWHM from the convolved arc lamps. Not doing this properly could result into inaccurate velocity dispersion determination. 

With the convolved and updated set of templates, the 63 Myr lowest metallicity template was no longer being fit and instead, as expected, there are contributions from templates $<$5 Myr at more reasonable, higher metallicities i.e. [M/H] = 0.0 and [M/H] = +0.22. 

Another factor that changes the line profile and isochrone, and therefore the integrated spectra for a stellar population at a given age and metallicity, is $\alpha$-enhancement. Alpha elements (O, Mg, Si, S, Ca, Ti) are dispersed into the interstellar medium by core-collapse supernovae as a result of massive star death, but are then diluted by supernova Type Ia (SNe Ia) events that do not contribute as much $\alpha$ elements. An $\alpha$-abundant stellar population has an isochrone track shifted to cooler temperatures (or redder colors). Early-type galaxies and galaxy regions that have undergone rapid star formation have been found to have enhanced $\alpha$-abundances, having very little successive star formation events to dilute them. We therefore investigated the contribution of $\alpha$-enhanced stellar populations in the different components of NGC 2903.

\begin{figure*}
\begin{center}
\includegraphics[width=\textwidth]{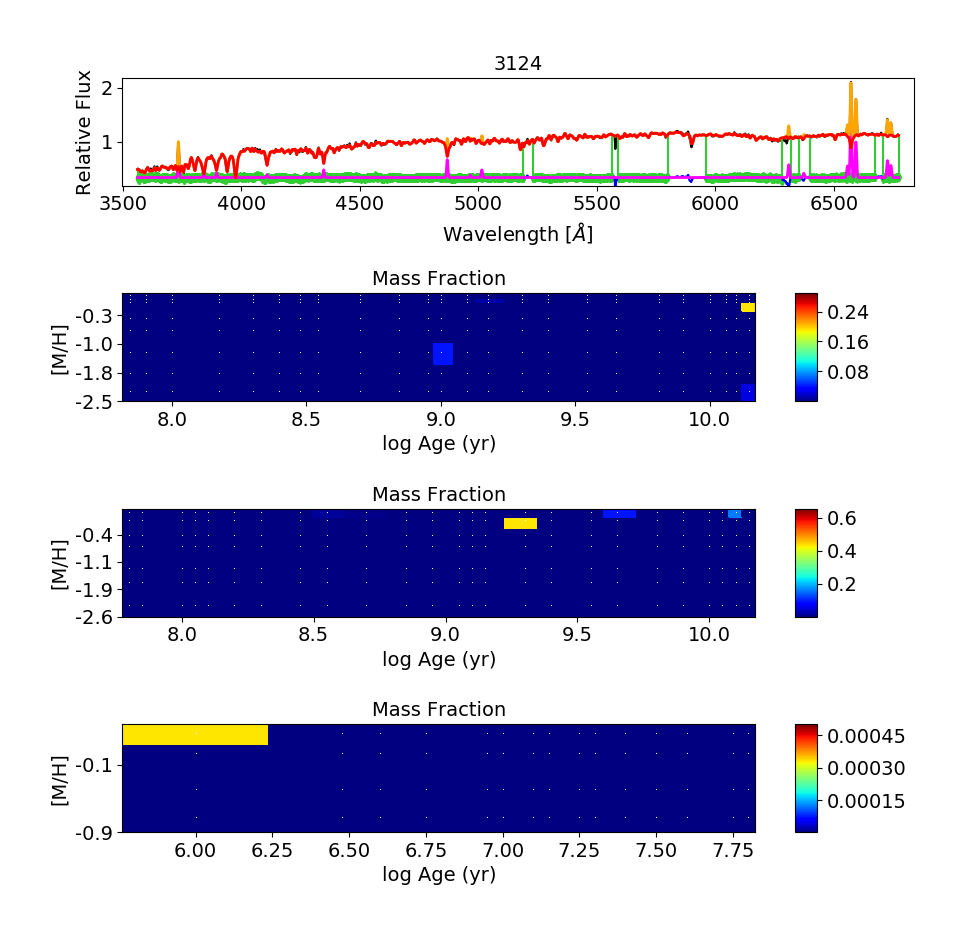}
\includegraphics[width=\textwidth]{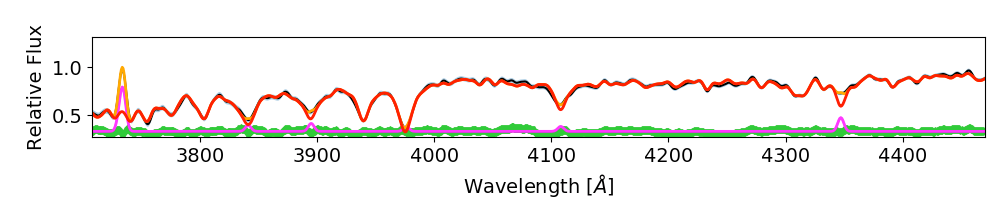}

\caption{pPXF output for spaxel 3124 in NGC 2903. Top panel: observed VENGA data (black) with the model for the continuum (red), model for the gas flux (pink), combined gas and continuum models (orange), the residuals(green), and wavelength ranges not included in the fit(blue). We zoom into a smaller portion of the spectra (bottom panel) to highlight how well the model spectra (orange) reproduces the observed spectra (black). Middle 3 panels: contributions from different stellar populations in a grid of [M/H] (dex) vs log Age (yr) for the MILES alpha-enhanced templates (second panel, from the top), MILES solar-abundance templates (third panel), and young GRANADA templates (fourth panel). Take note that each sets of templates have different heat scales for their contribution to the bestfit template. This figure highlights pPXF's ability to reproduce the observed spectra from a combination of templates that span a wide range in ages and metallicities.}
\label{fig:ppxf_sample}
\end{center}
\end{figure*}

The alpha-enhanced SSPs \citep{vazdekis15} use the same MILES empirical stellar library, Salpeter IMF, and BaSTI isochrones \citep{Pietrinferni04,Pietrinferni06} with $\alpha$ abundance, [$\alpha$/Fe] = +0.4. This is different from the scaled solar model in that for the scaled solar model, [Fe/H] = [M/H] while for the alpha-enhanced model, [Fe/H]~=~[M/H]~-~0.75[$\alpha$/Fe]. The templates obtained also have FWHM =~6~\AA~to match the other templates.  
See Figure \ref{fig:ppxf_sample} for a pPXF fit to a sample spaxel, employing the procedure outline in this section. 

\subsection{Treatment of dust}
\label{sec:dust}

In the previous section, we fit away the continuum in order to get the stellar populations based mostly on the presence and shapes of lines. However, in trying to get a mass for a spaxel and more importantly, a total mass of contributing spaxels to each galaxy component, we need to get the intrinsic, dust-corrected continuum light from the galaxy. 

We derived $E(B-V)$ from pPXF by fitting the best-fit spectra derived from the LP fit with a Calzetti curve \citep{calzetti94}. pPXF does this by determining how much the continuum shape has changed from an un-attenuated spectra i.e. the default for the template spectra. We then deredden the observed spectra, which we multiply by the M/L in the SDSS \textit{g}-band and corresponding weight derived from the LP fit, to get the \textit{g}-band mass for that spaxel for each contributing stellar population. We discuss this procedure in greater detail in Section \ref{sec:SFH}.

To check for consistency, we also derived the $E(B-V)$ from the Balmer decrement method. We apply Equation \ref{equation_} (adopted from \citealt{dominguez13}) using H$\alpha$ and H$\beta$ fluxes measured from pPXF to get the $E(B-V)$ for each spaxel, assuming H$\alpha$/H$\beta$ = 2.87 (Case B recombination at $10,000 K$ and $10^{2-4} cm^{-2}$, \citealt{osterbrock89}). Here, $k_{\lambda}$ is defined by the Calzetti curve. 

\begin{center}
\begin{equation}
E(B-V) = \cfrac{-2.5 log \bigg(\cfrac{[H\alpha/H\beta]_{obs}}{2.87}\bigg)}{k(\lambda_{H\alpha}) - k(\lambda_{H\beta})}
\label{equation_}
\end{equation}
\end{center}

\begin{figure*}
\begin{center}
\includegraphics[width=0.90\textwidth]{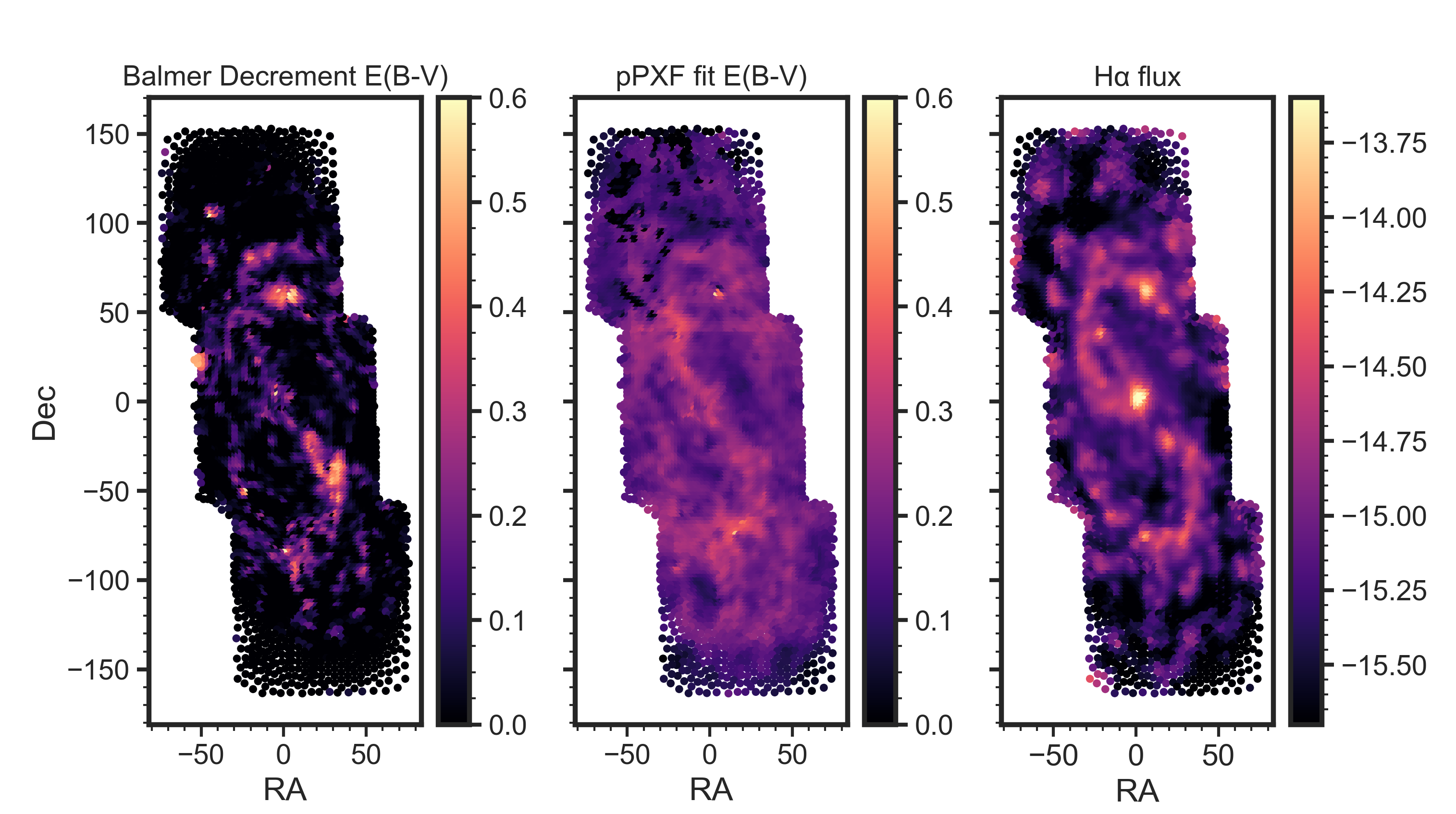}
\caption{$E(B-V)$ determined from Balmer decrement (left) using emission line flux from the gas, $E(B-V)$ determined from fitting the stellar continuum shape in pPXF (center), and non extinction-corrected H$\alpha$ flux ($\rm erg~s^{-1} cm^{-2}$ \AA$^{-1}$) map derived from the simultaneous gas-fitting with the stellar population-fitting in pPXF (right). We note that the areas with high extinction in both $E(B-V)$ maps trace similar regions i.e. areas with higher star formation as seen in the H$\alpha$ map. }
\label{fig:ebv}
\end{center}
\end{figure*}

However, one must be wary that the reddening derived from the gas is not necessarily the same as the reddening from the continuum. One complication to the Balmer decrement method is that H$\alpha$/H$\beta$ is a function of temperature and density and we assume that the emission is coming from a star-forming region. However, diffuse ionized gas (DIG) which has non-local origins, also contribute H$\alpha$ (see \citealt{kaplan16} for how they dealt with DIG emission in a sample of VENGA galaxies). 
Figure \ref{fig:ebv} shows the extinction maps derived from both methods as well as the non extinction-corrected H$\alpha$ map of NGC 2903. Both extinction maps show areas with higher values that correspond to where current star formation is happening as seen in the H$\alpha$ map and where the dust lanes are in the optical (Figure \ref{fig:sdss_image}). 
We use the reddening derived from the continuum to get the intrinsic light from the galaxy to be consistent with the stellar populations analysis.

\section{Mass Build-up and Star Formation History}
\label{sec:SFH}

The traditional way of determining mass is by using an infrared image, where the dominant light source are K giants. This justifies our light decomposition in Section \ref{sec:galfit} where we use the 2MASS $K_s$-band data in decomposing the galaxy into the bulge, bar, and disc. 
But to be more consistent with how we derived the stellar population properties from the template fitting, we compute the mass from the M/L ratio for each stellar population (of different metallicities) at different ages, by knowing how much they contribute to the total light. 

We use the M/L in the SDSS $g$ band since its whole range is within the observed wavelength of VENGA and within the wavelength range of our SSP templates. The MILES stellar populations \citep{vazdekis10} has magnitudes, colors, M/L, and masses tabulated in their website\footnote{http://miles.iac.es/} for all their single stellar populations in different filters, including SDSS. As for the GRANADA-based templates, we derived the M/L ourselves using the fact that the spectra are in units of $\rm L_{\odot}/M_{\odot}/$\AA. We convolve the spectra with the SDSS \textit{g}-band transmission curve and integrate.  
We also take into account that the contributing mass to the spectra is a combination of the stellar remnants and stars, and is smaller than the initial mass of the cloud (at 1 $\rm M_{\odot}$). Therefore the per $\rm M_{\odot}$ is actually interpreted as per $\rm M_{\odot}$ $\times $ remaining mass fraction left in remnants and stars. For example, this fraction is 1 i.e. no mass lost yet for a 10 Myr population and $\sim$ 0.7 for a 10 Gyr stellar population. 

To double-check the resulting M/L, we applied the same method to the MILES templates that already have the M/L in the $g$ band. They are at most, different by only 3\%, with our values being higher. For self-consistency, we use the derived M/L instead of the tabulated M/L for the MILES templates and warn the reader that the mass we compute may be overestimated by 3\%.

Once we have the M/L in the $g$ band for all the templates and the dust-corrected light from the galaxy as outlined in Section \ref{sec:dust}, we can then multiply them by each other and get the mass of the spaxel if 100\% of its light came from that stellar population. This has to be multiplied by the weights determined from the stellar population fitting and then added up to get the total mass of the spaxel in g-band. 

\begin{figure*}
\begin{center}
\includegraphics[width=0.45\textwidth]{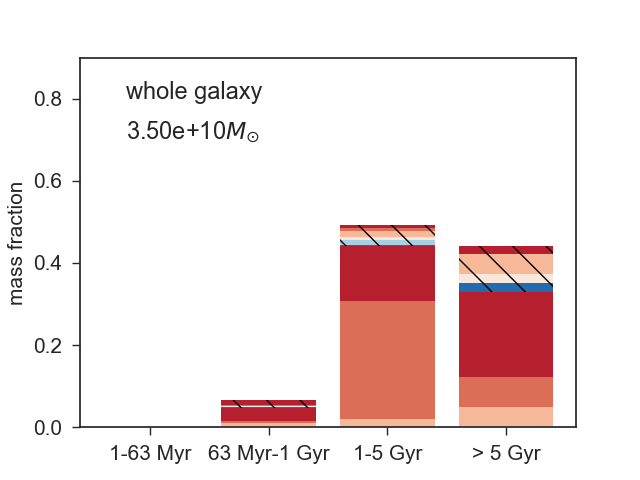}
\includegraphics[width=0.45\textwidth]{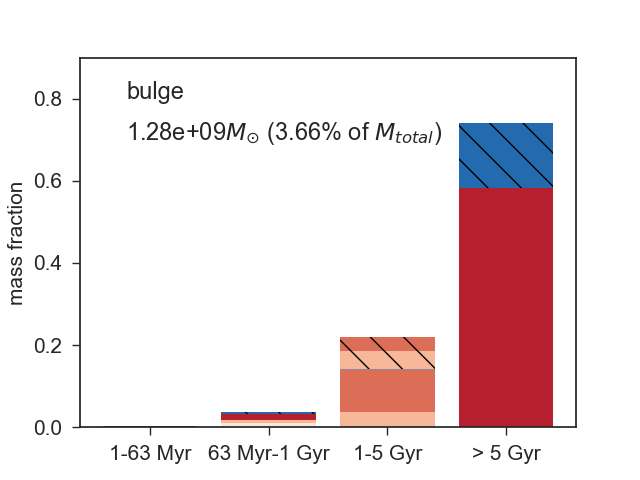}
\includegraphics[width=0.45\textwidth]{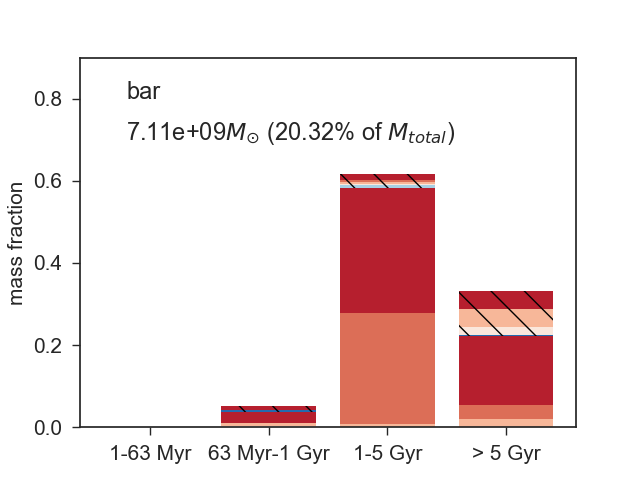}
\includegraphics[width=0.45\textwidth]{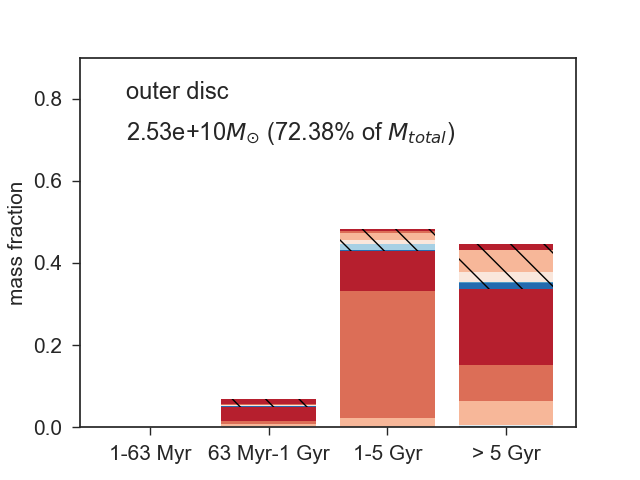}
\includegraphics[width=0.45\textwidth]{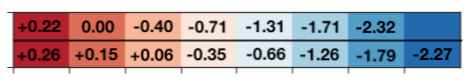}

\caption{Clockwise from top left: stellar  mass fraction for the whole galaxy, the bulge, the outer disc, and the bar, of NGC 2903 in 4 age bins, color-coded by the contribution of different metallicities. The vertical axis shows the mass fraction of the mass of the galaxy or galaxy component, indicated on the top left of each panel. The contribution of the alpha-enhanced templates are shown as the hashed part of the bar graphs. The color bar below the panels shows the corresponding metallicity ([M/H]) for the solar abundance templates (top) and the alpha-enhanced ([$\alpha$/Fe]=0.4) templates.} 
\label{fig:massbuildup}
\end{center}
\end{figure*}

Using the spaxel membership from Section \ref{sec:galfit}, we combine the masses in four different age bins for spaxels that belong to the same component, color-coded by the contribution of different metallicities. Figure \ref{fig:massbuildup} shows these mass fraction build up for the whole galaxy, the bulge, the bar, and the outer disc. We group the ages accordingly: 1-63 Myr for stellar populations coming from the GRANADA templates and whose light is dominated by current star formation, 63 Myr - 1 Gyr for the younger MILES templates that are also dominated by light from O and B stars, 1 - 5 Gyr for the stellar populations dominated by A type stars, and $>$ 5 Gyr for those dominated by F and G stars. \citet{conroy13} (Stellar Population Synthesis Review) expressed that SPS with full-spectrum fitting can be trust-worthy up to 7 Gyr, after which, it is harder to discern between older populations. However, we choose to be conservative with our age-binning based on the lifetime of A-type stars whose age we can tell from the Balmer lines and Balmer Break. 

 We derived masses for NGC 2903, listed in Table \ref{tab:masses}, from light-weighted and mass-weighted stellar population fitting. 
The light-weighted masses are lower which is expected, as younger stellar populations are weighed more in this case which have lower mass-to-light ratios. Because the total light-weighted mass of NGC 2903 is smaller than mass measurements in the literature (i.e. 21\% of mass from \citealt{lee09}) and the mass-weighted stellar population fitting from this study (22\%), we do not discuss these results in the context of the mass growth of each galaxy component  (Section \ref{sec: components}).   

\begin{table*}
\begin{center}
\caption{Total masses for each galactic component of NGC 2903 from mass-weighted and light-weighted stellar population fitting}
\label{tab:masses}
\begin{tabular}{c  c  c  c  c }
\hline
galaxy component & mass-weighted ($M_{\odot}$) & \% of total MW mass & light-weighed ($M_{\odot}$) & \% of total LW mass\\
\hline
whole galaxy & $3.50 \times 10^{10}$ & 100 & $7.70 \times 10^{9}$ & 100\\
bulge & $1.28 \times 10^{9}$ & 3.66 & $1.83 \times 10^{8}$ & 2.38 \\
bar & $7.11 \times 10^{9}$ & 20.32 & $1.62 \times 10^9$ & 21.04 \\
outer disc & $ 2.53 \times 10^{10}$ & 72.38 & $5.73 \times 10^{9}$ & 74.42\\
\hline
\end{tabular}
\end{center}
\end{table*}

\section{Results and Discussion}
\label{sec:results}

One major advantage of our spatially resolved IFU-based study is that we can dissect the galaxy into its outer disc, bar, and bulge, and analyze the stellar populations of each component. Figure \ref{fig:massbuildup} shows the age, metallicity, and mass of the stellar populations present in the  whole galaxy, as well as in its  outer disc, bar, and central bulge. The y-axis shows the stellar mass fraction as determined in Sections \ref{sec:sps} and \ref{sec:SFH}. The x-axis shows the four age bins used for the fits, as discussed in Section \ref{sec:SFH}.

The hashed marks show the contribution from alpha-enhanced stellar populations, and the color bar below the four panels show the metallicity ([M/H]) for the solar abundance templates  and the alpha-enhanced ([Mg/Fe]=0.4) templates. One can see that at smaller lookback times (i.e., for younger stars), the contribution from alpha-enhanced stellar populations decreases. This is consistent with chemical evolution models where the $\alpha$ elements produced in core-collapse supernovae are diluted by SNe Ia events.

We next discuss the properties of the stars in the outer disc, bar, and bulge of NGC 2903 and the implications for the assembly history of NGC 2903.

\subsection{Stellar population ages and metallicities in the outer disc, bar, and bulge}
\label{sec: components}

{\bf{The outer disc:}}  The bottom right panel of  Figure \ref{fig:massbuildup} shows the stellar populations of the outer disc of NGC 2903. The  outer disc contributes to  72\% of the total stellar mass of the whole galaxy,  16\% lower than the Disk/T from the GALFIT decomposition. This is because of the absolute nature of our spaxel assignment to only one component, when in fact there could be superposition of different components in a given region. Specifically, as seen in Figure \ref{fig:subcomponents}, the outer disc light distribution peaks in the central region but these spaxels are assigned to either the bulge or the bar in Figure \ref{fig:spaxels}. The outer disc has 46\% of its stellar mass in the form of stars of age $\rm > $ 5 Gyr,  48\% stars of age between 1 and 5 Gyr, and the remaining small fraction (below 10\%)  of its stellar mass is locked in young stars.

It is clear that the outer disc of NGC 2903 does not exhibit a simple exponentially declining star formation history as is sometimes assumed in doing SED fits of galaxies out to intermediate galaxies. Rather, this SFH is more complex. The observed properties support a scenario where a) nearly half of the stellar mass in the outer disc of massive spirals formed via recent (1-5 Gyr ago) in-situ star formation from gas that could have accreted recently or over an extended period of  time \citep{keres05,dekel09}; and b) the remaining substantial fraction of the outer disc may have been built via older ($>$5 Gyr) in-situ star formation as well as mergers of satellites \citep{naab06,khochfar09,bezanson09,jogee09,weinzirl09} that can have older ($\rm > $ 5 Gyr)  and/or metal-poor stars.
\\
\\
{\bf{The outer disc and stellar bar:}} As shown in the bottom left panel of Figure \ref{fig:massbuildup}), the stellar bar contributes to 20\% of the total stellar mass of the galaxy, and it is about 3 times less massive than the outer disc. We measure 14\% more than the Bar/T (0.06) in the GALFIT decomposition. This could be due to our relatively lenient criterion on spaxel assignment (i.e. component dominance by $>50$\%) as discussed in Section \ref{sec:galfit}. The outer disc and bar have 46\% and 35\% of their respective stellar masses made of stars of age $\rm > $ 5 Gyrs, and 48\% and 65\% made of stars of age  between 1 and 5 Gyr. For the most part, the metallicity make-up of stars in the outer disc and bar are quite similar.
The similarity is broadly consistent with the general picture that stellar bars form (spontaneously or via tidal triggers) from disc material \citep{miller68,combes81}.

We perform a more detailed exploration of what one might expect by looking at  cosmological simulations  of bar formation by \citet{romano08}. These simulations show that at high redshifts (e.g., z $\sim$ 5-8),  the bars that form are transient in nature, repeatedly getting destroyed and reforming. The fragility and transient nature of the early generations of bars at these epochs are due, at least in part, to the fact that the outer disc is not yet very massive and is experiencing  frequent galaxy mergers, associated rapid changes in the gravitational potential, and large gas inflows. According to these simulations \citep{romano08}, stellar bars become more stable at z $\sim$ 1 (lookback times of about 8 Gyr), when the outer
disc has also become more massive and stable.

The fact that the bar in NGC 2903 has a larger fraction of its mass (65\%) in younger (1-5 Gyr old) stars than the outer disc, and has very few metal-poor stars is consistent with the above picture, and with the ability of bars to drive gas inflows. Specifically, a bar that forms at lookback times of 8 Gyr or less from a massive stellar disc would include only disc stars inside the corotation radius (CR) of the bar. These stars would therefore be from the inner parts of the outer disc, rather than in its outskirts, and would consequently tend to be metal rich than disc stars. Another factor is that the bar can very efficiently drive gas from the outer disc into the central kpc  by driving gas inflows on its leading edge \citep{shlosman93,knapen95,athanassoula03}. If the amount of shear along the bar is not too strong, the inflowing gas can form stars along the bar, thus adding young stars in the bar region \citep{jogee05}.
\\
\\
{\bf{The bulge and outer disc:}} As shown in the top right panel of Figure \ref{fig:massbuildup}, the central bulge has 4\%  of the total stellar mass of the galaxy.This is comparable to what GALFIT gives (5.7\%). Most (75\%) of the stars in the bulge are older than 5  Gyr,  and it has a sharply declining fraction of younger stars. This star formation history is reminiscent of that of early type galaxies. The presence of a small fraction (below 10\%) of young stars (63 Myr to 1 Gyr old) is consistent with gas inflow driven by the bar and other mechanisms building a younger stellar population in the central  part of the galaxy.  The presence of old, high metallicity stars is intriguing and will be  discussed in $\S$~\ref{sec: bulge}.
 
It is striking that  while the bulge has 75\% of its stars with age $\rm >$ 5  Gyr, the  outer disc has over 50\% of its stars with younger ages  $<5$ Gyr. The chemical evolution of the bulge is also different from the bar and the disc, with a large fraction of low metallicity, high-$\alpha$ stars at $\rm >$ 5 Gyrs, similar to what we see for the alpha-metallicity trends in the Milky Way bulge \citep{ness16, bensby17}. These properties of the bulge are consistent with aspects of the inside-out growth of galaxies as they support the idea that older and/or metal-poor stars form first in the central regions of galaxies, while younger stars form or assemble in the outer regions later.
\\
\\
Our results also agree with the structural study by \citet{weinzirl09} which suggests, based on the comparison of a large sample of spirals with predictions from semi-analytical models \citep{khochfar05, khochfar06, hopkins09}, that spiral galaxies with low B/T $\rm < 0.2$, such as NGC 2903 (which has B/T $\sim$ 0.057 from our GALFIT decomposition), have not undergone a major merger since z$\sim$2 (lookback time of $\sim$ 10 Gyr). Instead, such spirals have likely  built their outer discs as well as part of their bulge at later times through gas accretion, minor mergers, and secular processes.

More recently, there have been studies with hydrodynamical simulations of galaxies that focus on disc and bulge formation. \citet{gargiulo19} used the Auriga simulations \citep{grand17} to look at galaxies with S\'ersic index n $<$ 2 and $0.12 \rm <$  B/T $\rm < 0.44$. Comparisons with their galaxies that have $\sim$20\% of the bulge particles at ages between 1-5 Gyr (as we see in Figure \ref{fig:massbuildup} for the bulge) show low B/T (i.e. $\rm < 0.2$) and growth through minor mergers and secular processes. \citet{tacchella19} used IllustrisTNG hydrodynamical simulations \citep{pillepich18} to study the spheroidal and disc components of galaxies. They find that similar-mass galaxies to NGC 2903 have bulges that grow very little between the epoch of disc formation (z$\sim$2) and z=0, which therefore decreases the B/T from $\sim$0.45 to $\sim$0.30. Though these B/T estimates are higher than what we find for NGC 2903, we want to highlight that this decrease is due to the disc mass build-up and lack of major mergers during the same epoch. 

\subsection{2D maps of stellar age and metallicity}
\label{sec:gradients}

\begin{figure*}
\begin{center}
\includegraphics[width=0.35\textwidth]{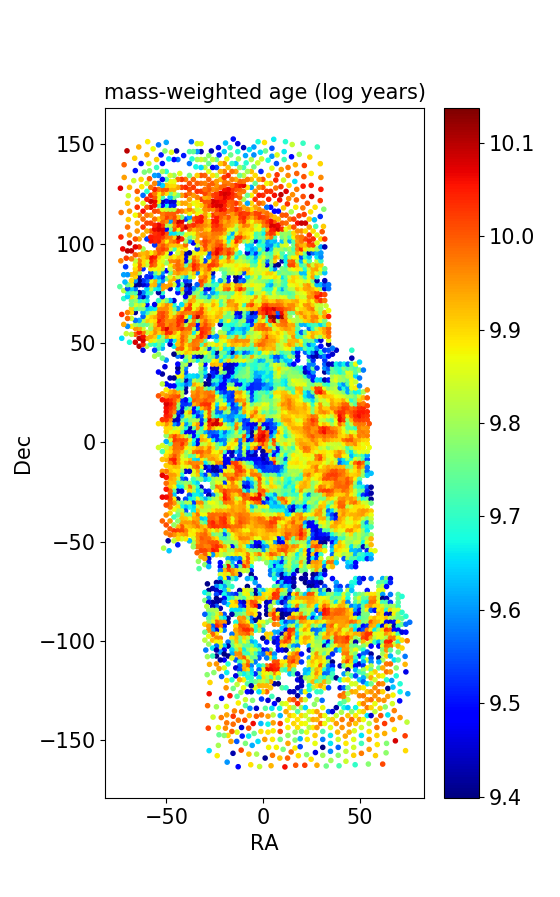}
\includegraphics[width=0.35\textwidth]{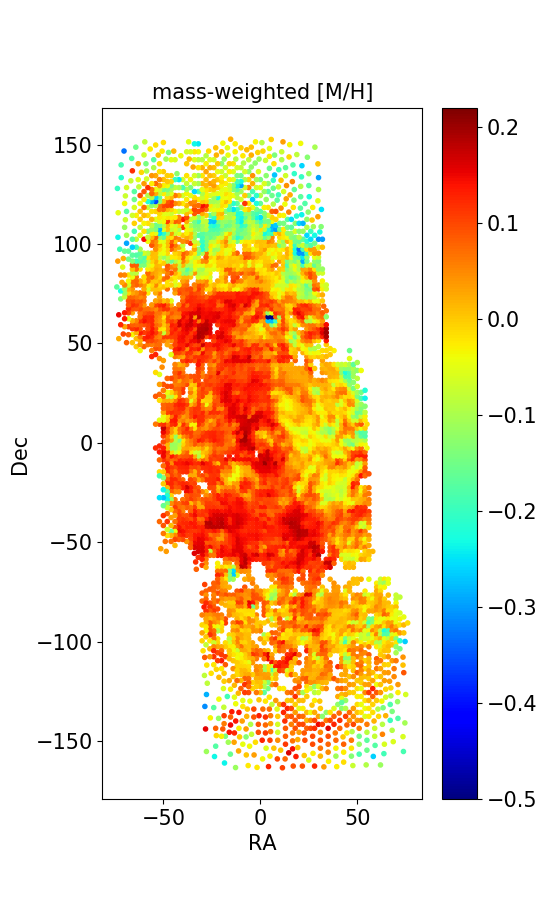}
\includegraphics[width=0.35\textwidth]{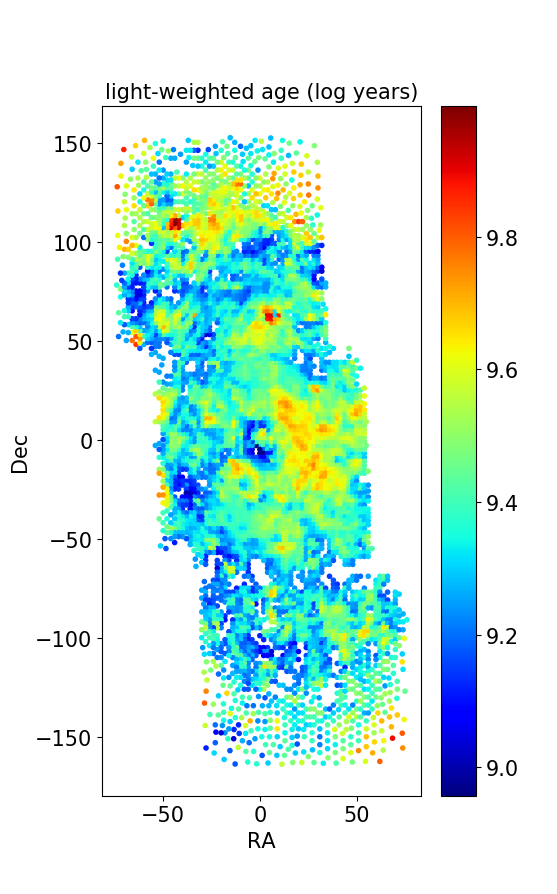}
\includegraphics[width=0.35\textwidth]{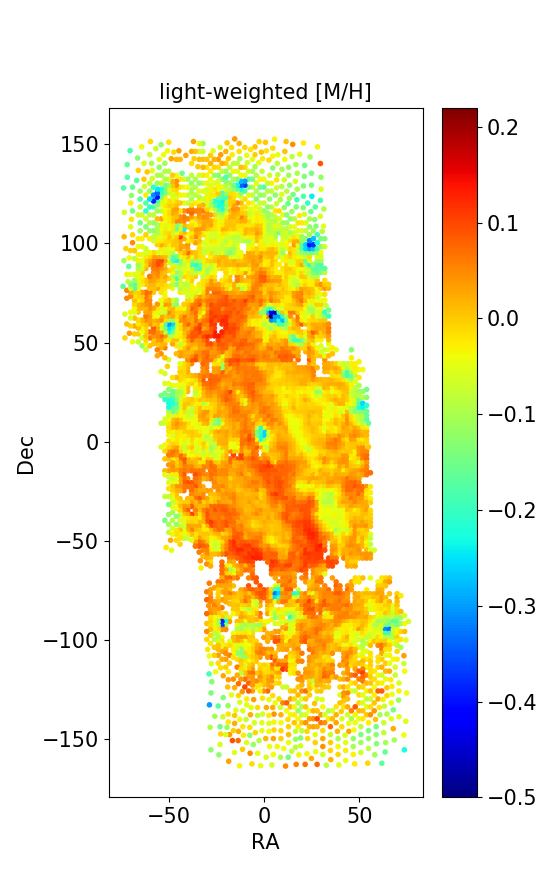}
\caption{Top: Mass-weighted average age (left) and average metallicity (right). The color bars represent log of age in years and [M/H], respectively. Bottom: Same as above but light-weighted. The light-weighted age and metallicity are younger and lower respectively, as these stellar populations have lower mass-to-light ratio compared to older and higher metallicity stellar populations. Both age maps show that the disc generally has no features but that the bulge has a young population and an old population. On the other hand, both the metallicity maps trace areas of high star formation (in the spiral arms and central regions), showing enhanced metallicities.  }
\label{fig: MWLW_maps}
\end{center}
\end{figure*}

From the stellar population synthesis and best-fit combination of stellar populations for each spaxel, we derived mass-weighted and light-weighted age and metallicity maps as shown in Figure \ref{fig: MWLW_maps}. Both the age and metallicity are higher for the mass-weighted maps than the light-weighted maps. This physically makes sense as older stars at a given metallicity and higher-metallicity stars at a given age have higher mass-to-light ratios and would therefore dominate the mass-weighted maps. On the other hand, the light-weighted maps show lower values for the average age and metallicity for a spaxel. Younger stars dominate the light for a given stellar population, therefore the light-weighted age is biased towards younger ages. Also quite noticeable is the enhanced metallicity, both in the mass-weighted and light-weighted maps, for the spiral arms and central region, areas with higher star formation. 

\begin{figure*}
\begin{center}
\includegraphics[width=0.32\textwidth]{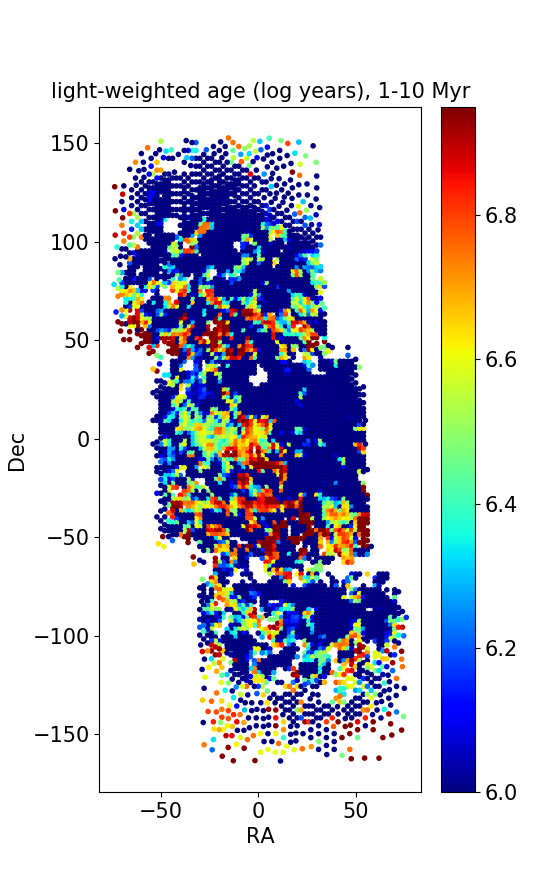}
\includegraphics[width=0.32\textwidth]{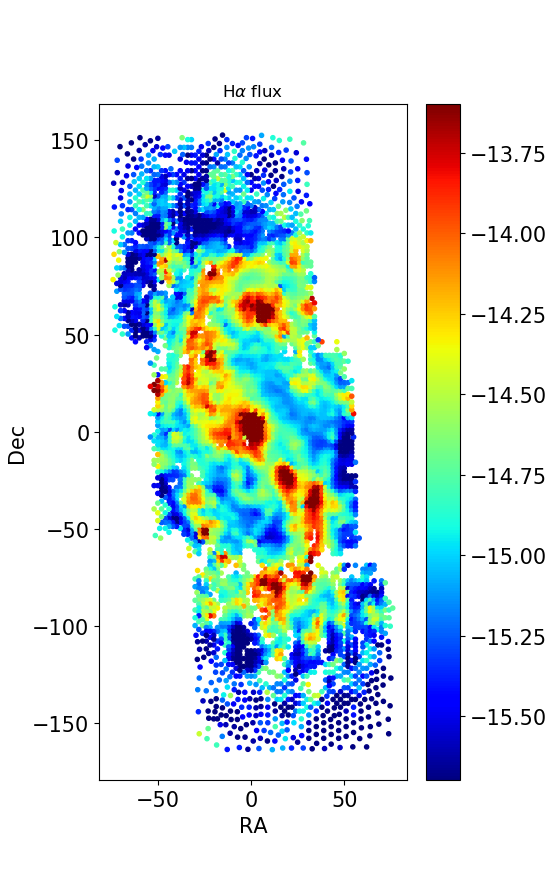}
\includegraphics[width=0.32\textwidth]{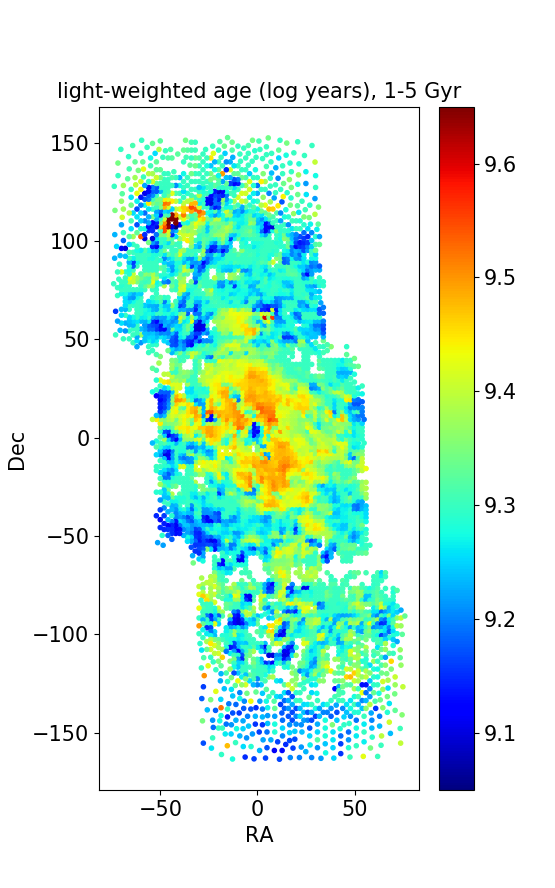}
\caption{Light-weighted average age for stars that are less than 10 Myr old (left), extinction-corrected H$\alpha$ map (center), and light-weighted average age for stars with ages between 1 and 5 Gyr (right), to compare areas with recent star formation based on the continuum and gas vs an older population. The color bars for the first and third panels represent log of age in years while the color bar for the second panel is log of the extinction-corrected H$\alpha$ flux in erg$\rm s^{-1} cm^{-2}$ \AA$^{-1}$. H$\alpha$ recombination lines trace stars formed $<$ 10 Myr. As expected, the young stellar population map agrees with the features being traced in the H$\alpha$ map. The stellar populations with ages 1-5 Gyr do not have the same features as the H$\alpha$ map and instead shows older ages inside corotation i.e. location of the bar. }
\label{fig: maps_ages}
\end{center}
\end{figure*}

To look into these patterns further, we made light-weighted averaged age maps for stellar populations with ages $\rm < 10$ Myr. We also compare it to the H$\alpha$ map derived from the simultaneous gas-fitting of pPXF together with the stellar population fitting. These are shown in the left and center panels of Figure \ref{fig: maps_ages}. For the map of younger stars ($\rm < 10$ Myr), the fact that the bulge and the spiral arms have relatively older ages means that there is a continuous production of stars in those regions for the last 10 Myr, as seen and traced by the H$\alpha$ map. We compare these two maps to intermediate-age stars (1-5 Gyr) as seen on the right panel of Figure \ref{fig: maps_ages}. This map, on the other hand, does not share the same features as the H$\alpha$ map. 


It is also interesting that we see two populations in the central region, though we note that only the innermost spaxels were classified as part of the bulge in making the mass build up histories. From Figure \ref{fig: MWLW_maps}, we can see a younger and higher metallicity stellar population surrounding an older, lower metallicity stellar population in both the light-weighted and mass-weighted maps. This is evidence of the co-existence of an older and lower metallicity bulge and a younger and higher metallicity bulge. Such composite systems have been found in other galaxies through light profile and kinematics studies (e.g., \citealt{fisher10,erwin15}), preferentially in spirals with bars. In this study, we distinguish these composite bulges in metallicity and age, complementary to previous work.   
The presence of younger stars in the bulge as well as our structural fit that gives it a S\'ersric index of n $<$ 2 hints that we may be seeing NGC 2903's disky pseudobulge \citep{kormendy04,fisher08}, though a more detailed study on its kinematics is needed.  

The presence of younger stars in the central region is also in line with low S\'ersric index and low B/T galaxies that have grown through secular processes, gas accretion, and minor mergers in the recent past \citep{weinzirl09}. For NGC 2903's case, we see this evident in the bulge, which includes new stars, partly formed from existing or accreted gas driven into the central regions by the stellar bar. The light-weighted age map in Figure \ref{fig: MWLW_maps} shows that these stellar populations have ages $\rm < 1$ Gyr compared to 2.5 - 4 Gyr for the central part of the bulge. The mass-weighted age map in Figure \ref{fig: MWLW_maps} shows an even bigger difference in the average age, with the younger population being $\rm < 2.5$ Gyr and the older population $\rm > 10$ Gyr.


\subsection{Stars in the bulge}
\label{sec: bulge}

Figure \ref{fig:massbuildup} shows the stellar mass fraction of the whole galaxy, bulge, bar, and disc, divided into bins of different stellar ages.  Most (75\%) of the stars in the bulge are older than 5 Gyr  and largely have high metallicity, followed by low metallicity and alpha-enhanced compositions. In this section, we try to understand the metallicity pattern better and explore whether it is plausible to have such high metallicities in bulge  stars with age $\rm >$ 5 Gyr.

We first compare to observations of gas-phase metallicity radial gradients at high redshifts. These observations were first done using lensed systems \citep{jones10,yuan11} where it is easier to get spatially-resolved measurements. More observational studies \citep{swinbank12,jones13,molina17,wang17} have been made possible, both for lensed and unlensed systems, with powerful IFS and adapative optics. These studies find a range of slopes and trends for the gas-phase metallicity gradient, with \citet{swinbank12} finding two high redshift galaxies exhibiting super-solar metallicities at their central regions.These observations suggest that old, high metallicity stars could be built from this gas in the central regions of galaxies at lookback times as early as 10 Gyr ago. There is hence  a plausible pathway for explaining the presence of bulge  stars with age $\rm >$ 5 Gyr and  super-solar metallicities in NGC 2903.

Next, we compare to observations of gas-phase metallicity radial gradients in the local universe. NGC 2903 is one of the galaxies that \citet{kaplan16} studied. The authors found that seven out of the seven gas-phase metallicity indicators exhibit solar and super-solar values in the central region of NGC 2903, even reaching 0.4 dex above solar values. It is interesting to note that other radial gradient studies in the low redshift regime (e.g., \citealt{gonzalezdelgado15,scott17}) show that the median stellar metallicity trend for galaxies with similar morphology and similar mass to NGC 2903 have slighty below solar stellar metallicity values. This may however be an effect of lower spatial sampling that would drive the average metallicity down. 

We also compare our findings on the NGC 2903 bulge to that of the Milky Way bulge. A complete study of the stellar population in the Milky Way is hampered by a lot of systematic uncertainties and observational constraints. We simply do not have well-characterized spectra for the short-lived stellar evolutionary phases to get a complete census of the mass fraction of differently-aged stars. However, it has been shown that the Milky Way bulge has had rapid star formation and chemical enrichment at or before z$\sim$2 and so a large part of its mass would be from those very old metal-rich stars \citep{ness13}.

\section{Conclusions}
\label{sec:conclusion}

In this study, we have used the VENGA IFU survey to dissect the age, metallicity, and mass of stars in the bulge, bar and outer disc of NGC 2903, and how their ensemble formation histories paint a comprehensive picture of galaxy evolution. We performed bulge-bar-disc decomposition with GALFIT to better assign a spectra to a galaxy component and performed full-spectrum fitting with pPXF to derive the stellar populations from which we got star formation histories. Our main results are:

(1) {\bf{The outer disc}}:  The outer disc has 46\% of its stellar mass in the form of stars with ages $\rm > $ 5 Gyr,  48\% in stars with ages between 1 and 5 Gyr, and $<$ 10\% in younger stars. There is a larger fraction of solar and sub-solar metallicity stellar populations in the outer part of the disc compared  to the inner parts of the galaxy.  These observed  properties support a scenario where the outer disc may have mainly formed via  in-situ star formation recently (1-5 Gyr ago) and at earlier times ($>$ 5 Gyr ago), and via mergers of satellites with older (e.g. age  $\rm > $ 5Gyr) and/or metal-poor stars.

(2)  {\bf{The stellar bar}}: The stellar bar is about three times less massive than the outer disc. It has a broadly similar stellar metallicity make-up and distribution of stellar ages as the outer disc, but has a higher fraction of younger stars of ages between 1 and 5 Gyr (65\%  in the bar vs 48\% in the outer disc). These results support the general picture where stellar bars form, spontaneously or via tidal triggers, from disc material, and where gas inflows along a weak or moderately strong bar can lead to star formation along the bar.

3) {\bf{The bulge}}:
The central bulge contributes less than 5\%  of the total stellar mass of the galaxy and bulge-bar-disc decomposition of the  2MASS Ks-band image shows it has a low S\'ersic index of $\sim$ 0.27 (see Table \ref{tab:galfit_decomp}), suggestive of a disky pseudobulge rather than a classical bulge with a de Vaucouleurs profile. Its distribution of stellar ages and metallicity differs strongly from that of the bar and the outer disc. The bulge has 75\% of its stars with age $\rm >$ 5  Gyr,  while the bar and the outer disc have over 50\% of its stars with ages between 1 and 5 Gyr or younger. These old  $\rm >$ 5 Gyr bulge stars largely have high metallicity, followed by low metallicity and alpha-enhanced compositions, similar to what we see for the alpha-metallicity trends in the Milky Way bulge. These properties are broadly consistent with the idea of inside-out growth in the sense that the dominant part of the bulge likely formed earlier than the outer disc from central star formation over 5 Gyr ago.  The presence of a small fraction (below 10\%) of young stars (63 Myr  to 1 Gyr old) in the bulge is consistent with gas inflow driven by the bar and other mechanisms, building a younger disky bulge superimposed onto an older bulge.

(4)  We compare 2D mass-weighted and light-weighted stellar metallicity and age maps and find, as expected, that light-weighted maps are biased towards younger stellar populations that dominate the light. There is enhanced metallicity in the spiral arms and the central regions, both of which have higher star formation rate density as shown in the H$\alpha$ map.

(5)  Our results  in (1) to (3) are consistent with the study by \citet{weinzirl09}, which suggests that  most spiral galaxies in  the local Universe with low B/T have not had a major merger since  $z\sim 2$, and have grown their outer disc mostly through gas 
accretion, minor mergers, and secular processes,  based on their comparison of structural properties of spirals  (such as the distributions of  S\'ersic index  and $B/T$ of their bulges) to semi-analytic models of galaxy evolution from \citet{hopkins09}, \citet{khochfar05} and \citet{khochfar06}. This is further supported by recent hydrodynamical simulations (e.g.,
\citealt{gargiulo19,tacchella19}) that show that galaxies with B/T, S\'ersic index, and mass similar to NGC 2903 have not experienced
a major merger in the recent past.

In this work, we have emphasized the importance of the comprehensive study of individual and synergistic stellar populations and masses of the different parts of a galaxy to fully understand its formation and evolution. With more expansive spectral templates, and telescopes like JWST and the Giant Magellan Telescope (GMT\footnote{https://www.gmto.org/}) that let us peer deeper into space at higher resolutions, we will be able fill in our knowledge about how galaxies and their components change over time. Doing it in the nearby universe as we did in this study is but the first step.

\section*{Acknowledgements}

This paper includes data taken at The McDonald Observatory of The University of Texas at Austin. A.C. would like to thank Keith Hawkins, Milos Milosavljevic, Jason Jaacks, and Rodrigo Luger for insightful discussions and Rosa Gonzalez-Delgado for providing the Base GM templates. A.C. and S.J. acknowledge support from National Science Foundation (NSF) grant AST-1413652, NSF AST-1614798, NSF AST-1757983 and the McDonald Observatory and the Department of Astronomy Board of Visitors. A.C. also thanks the Large Synoptic Survey Telescope Corporation (LSSTC) Data Science Fellowship Program, which is funded by LSSTC, NSF Cybertraining Grant 1829740, the Brinson Foundation, and the Moore Foundation.






\bibliographystyle{mnras}
\bibliography{venga}




\bsp	
\label{lastpage}
\end{document}